\documentclass[aps,physrev,preprint,groupedaddress]{revtex4-2}

\usepackage{graphicx}
\usepackage{dcolumn}
\usepackage{bm}
\usepackage{booktabs}
\usepackage{subcaption}
\begin{document}
\preprint{APS/123-QED}

\title{\textbf{Three-Dimensional Simulation of the University of Hawai‘i FEL Oscillator with Cavity Desynchronization}}

\author{Amir Weinberg}

\affiliation{Department of Physics and Astronomy, University of Hawai`i at Mānoa, Honolulu, HI 96822, USA}
\author{Levi K.C. Fisher}
\affiliation{Department of Physics and Astronomy, University of Hawai`i at Mānoa, Honolulu, HI 96822, USA}
\author{Eremey Valetov}
\affiliation{Department of Physics and Astronomy, University of Hawai`i at Mānoa, Honolulu, HI 96822, USA}
\author{Siqi Li}
\email{siqili@hawaii.edu}
\affiliation{Department of Physics and Astronomy, University of Hawai`i at Mānoa, Honolulu, HI 96822, USA}

\date{\today}

\begin{abstract}
In this paper, we present three-dimensional, time-dependent simulations of the University of Hawai‘i (UH) at Mānoa free-electron laser (FEL) oscillator. 
Using beam parameters from the UH facility, we study the pass-by-pass evolution of the radiation field, including its temporal, spectral, and transverse properties. 
At nominal bunch length, the radiation pulse develops temporal spiking near saturation, together with sideband formation and increased sensitivity to machine timing jitter.
Our results show that modest cavity desynchronization can enhance the radiation energy by 63\%. Large cavity desynchronization, on the other hand, can effectively suppress the spiking instabilities and improve robustness to timing fluctuations.
Finally, we simulate a short-bunch operational mode with a bunch length comparable to the slippage length, which accelerates saturation and further amplifies the FEL power. 
Overall, these results provide a quantitative foundation for pulse control studies in the UH FEL oscillator and a critical benchmark for future experimental validation and machine optimization.
\end{abstract}

\maketitle

\section{\label{sec:intro}Introduction}
Free-electron lasers (FELs) are powerful light sources capable of generating high-power, tunable radiation spanning wavelengths from the infrared (IR) to the x-ray regime. Among different types of FEL configurations, oscillator-based systems offer unique advantages, including phase-stable pulse trains and narrow spectral bandwidth, where the radiation pulse interacts with a fresh electron bunch after each roundtrip in the oscillator cavity. These features make FEL oscillators attractive for applications such as precision spectroscopy and coherent spectroscopy \cite{ramian1992new, oepts1995free, winnerl2006felbe, schollkopf2013ir, zen2016present}.

The dynamics of FEL oscillators differ significantly from those of single-pass amplifiers due to the cavity feedback mechanism and the multi-pass evolution of the optical pulse. Understanding phenomena such as superradiant emission, cavity desynchronization, and saturation dynamics in oscillators requires detailed, time-dependent modeling. Recent advances in generating extremely short, high-power x-ray FEL pulses~\cite{franz2024terawatt} have motivated theoretical development and three-dimensional (3D) simulation efforts to investigate soliton-like superradiant behavior in amplifier FELs in saturation~\cite{robles2024three}. In FEL oscillators, a similar spiking behavior was observed in early experiments operating in the low-gain, highly saturated regime~\cite{warren1986spiking,richman1989first}. Recent interest in short-bunch superradiance for IR FEL oscillators also highlights the need for high-fidelity 3D simulation tools that can accurately model the interactions between the radiation and the electron bunch over many cavity round-trips~\cite{zen2023full}.

While simulation codes such as \texttt{GENESIS}~\cite{reiche1999genesis}, \texttt{GINGER}~\cite{fawley2006enhanced}, and \texttt{PUFFIN}~\cite{campbell2012puffin} have been widely used for FEL modeling, they each present limitations for oscillator studies. 
\texttt{GENESIS} and \texttt{PUFFIN} employ fully 3D field solvers, whereas the original \texttt{GINGER} assumed axisymmetric fields. \texttt{PUFFIN} can model strong gain and harmonic emission from first principles, but is computationally intensive and has not been widely applied to oscillator systems until recently~\cite{pongchalee2024unaveraged}. 
In addition, the \texttt{MINERVA} code, in combination with the Optical Propagation Code (OPC)~\cite{karssenberg2006modeling}, has been applied to oscillator studies, providing a three-dimensional, time-dependent framework that includes optical propagation in the cavity and can treat both high-Q and low-Q regimes~\cite{van2021three}. These developments highlight the growing ability of modern FEL simulation tools to handle oscillator physics, but also the continuing need for flexible infrastructures that can be adapted to different facilities and experimental configurations.

In this paper, we present a 3D simulation framework for FEL oscillators using a developmental version of the new \texttt{GINGER-3D} code, a major revision of the original \texttt{GINGER} code~\cite{fawley2006enhanced}, including a fully Cartesian x-y Alternating-Direction-Implicit radiation field solver in the transverse plane. We have developed an external pulse propagation module in \texttt{Matlab} and \texttt{Python} that couples the near-field radiation output of \texttt{GINGER-3D} and models cavity geometry and mirror optics, providing flexibility for implementing advanced features such as inserting additional optics for interferometric configurations to achieve phase coherence~\cite{oepts1992induced,szarmes2002michelson1,szarmes2002michelson2}.

As a first demonstration, we apply this simulation framework to the FEL oscillator with baseline parameters from the University of Hawai`i (UH) at Mānoa’s FEL facility. Originally developed by John Madey, the facility was designed with the capability of generating IR radiation via FEL interaction and x-ray radiation via inverse Compton scattering of the electron beam and the IR photons~\cite{niknejadi2019free}. Although it has been inactive for several years, current efforts focus on re-commissioning both the accelerator and the FEL, alongside supporting simulation and design studies~\cite{Bidault2025ipac, Weinberg2025ipac}.

Using the nominal parameters of the UH FEL, our simulations show that near saturation the radiation develops temporal spiking and spectral sidebands. While these are known features of FEL oscillator dynamics~\cite{warren1986spiking,richman1989first}, the present work resolves their full three-dimensional time-dependent evolution for the UH FEL oscillator and connects them to tuning behavior, including cavity desynchronization, transverse mode evolution, spectral structure, and radiation energy optimization.

We also investigate the effect of cavity desynchronization. A cavity is considered perfectly synchronized when the round-trip time of the radiation pulse matches an integer multiple of the electron bunch repetition period. However, during the FEL interaction, the radiation pulse slips ahead of the electron bunch, and the amplification process effectively slows the group velocity of the radiation pulse compared to its speed in vacuum. In the context of FEL oscillators, this is known as the laser lethargy effect~\cite{colson1990free}. By slightly shortening the cavity length, i.e. introducing desynchronization, the interaction between the radiation and the electron bunch can be sustained throughout the gain process~\cite{colson1983pulse}. At saturation, the group velocity of the radiation catches up as the gain slows down, suggesting that dynamic desynchronization would be ideal to optimize both gain buildup and saturated power~\cite{bakker1993dynamic}. This has previously been achieved by modulating the electron bunch repetition rate within a macropulse using fast RF phase shifters, without physically changing the cavity length~\cite{bakker1993dynamic,zen2020high}. 
However, such phase shifters are not currently available at the UH Mānoa facility. 
Therefore, in our simulation study, we implement desynchronization by shortening the cavity length by a fixed amount. 

In our simulations of the UH FEL oscillator, small cavity desynchronization, corresponding to a round-trip shortening by a fraction of the radiation wavelength, can enhance both the peak power and radiation pulse energy. At larger desynchronization, corresponding to a round-trip shortening of several radiation wavelengths, the main effect is to suppress the spiking dynamics that arise from the electron–radiation interaction near saturation. In this regime, temporal spiking instabilities and spectral sidebands are significantly reduced, producing cleaner and more stable pulse evolution, although with lower radiation power, consistent with the predictions in~\cite{kim1991stability}. We further show that large desynchronization makes the FEL oscillator more robust against timing fluctuations. In addition, the full three-dimensional model allows us to investigate the role of optical guiding in the pulse evolution.


To examine pulse evolution beyond the nominal UH FEL operating regime, we also consider a short-bunch case in which the electron bunch length is reduced to the sub-picosecond regime, comparable to the slippage length. This regime is of interest because shorter bunches can increase the peak current and therefore enhance the FEL radiation power, which would in turn improve the photon flux available for the downstream inverse Compton scattering x-ray source~\cite{niknejadi2019free}.
However, preliminary beam-dynamics studies indicate that compressing the nominal 2~ps bunch to the sub-picosecond regime is not achievable with the current machine configuration. Such compression would introduce increased slice energy spread, a correlated energy variation across the bunch, and significant beam transmission loss from the gun to the FEL entrance. The resulting energy chirp would also exceed the tolerance of the present untapered undulator. Despite this limitation, the short-bunch regime remains useful for understanding the oscillator physics and for guiding future upgrades. Beam manipulation and shaping techniques~\cite{cesar2021electron} may enable this operating mode and support applications such as IR absorption and vibrational spectroscopy with sub-picosecond resolution~\cite{kiessling2018synchronized}. Preliminary beam-dynamics studies of the UH beamline are being carried out using COSY INFINITY~\cite{COSYCAP04,berz2026cosy}, RF-Track~\cite{Latina:2016rftrack}, and Xsuite~\cite{iadarola2023xsuite}, and will be reported in future work. In the simulations presented here, we therefore vary only the bunch length while keeping the other machine and beam parameters fixed at their nominal values.


Our simulations show that the short-bunch case reaches saturation earlier and achieves higher output power than the nominal bunch-length case. The cavity desynchronization scan shows that modest desynchronization can further enhance the radiation energy, while the stabilization effect at large desynchronization and the transverse mode evolution differ from the 2~ps case because the bunch length is comparable to the slippage length.


The paper is organized as follows. Section~\ref{sec:sims_setup} describes the simulation tool and the beam and cavity parameters at the UH Mānoa FEL facility and demonstrates the simulation results of the nominal bunch operation. Section~\ref{sec:desync} presents the simulation results with nominal bunch length while implementing cavity desynchronization. Section~\ref{sec:shortbunch} presents the simulation results of the short-bunch operation along with the effects of cavity desynchronization. Section~\ref{sec:conclusions} offers concluding remarks and future research directions.

\section{\label{sec:sims_setup} The UH FEL Oscillator Simulation}
\begin{figure*}[!tbh]
   \centering
   \includegraphics*[width=\textwidth]{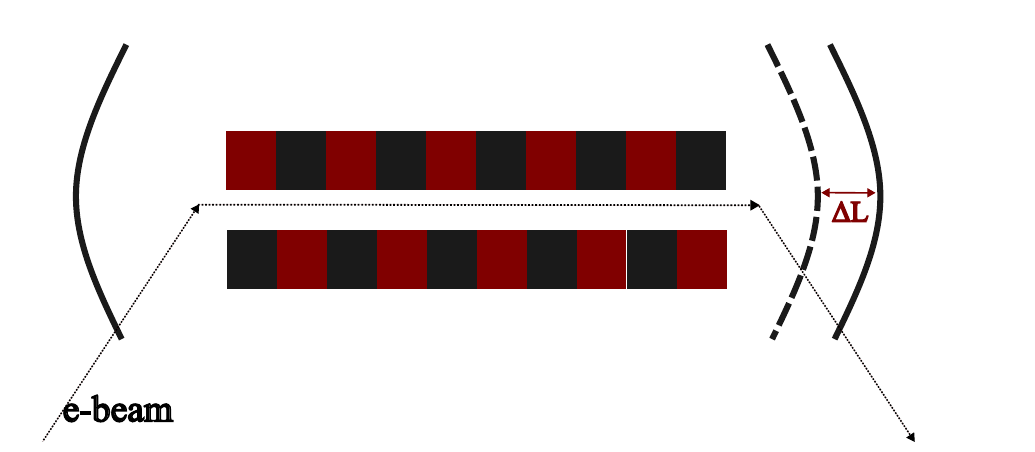}
   \caption{Schematic of the FEL cavity. $\Delta L$ indicates the shortening of the cavity length to implement cavity desynchronization.}
   \label{fig:schematics}
\end{figure*}

\begin{table}[!hbt]
   \centering
   \caption{Parameters of the UH Mānoa FEL.}
   \begin{tabular}{lcc}
       \toprule
       \textbf{Parameter} & \textbf{Value}                      & \textbf{Units} \\
       \midrule
        \textbf{e-beam} &&   \\ 
        Beam energy    & 40       & MeV     \\ 
        Energy spread    & 0.5      &     $\%$ \\ 
        Normalized emittance $\epsilon_{n,x}$, $\epsilon_{n,y}$      & 8        & $\pi$-mm-mrad  \\ %
        Bunch duration FWHM & 2 (nominal), 0.5 (short-bunch) &ps\\
        Current Peak     & 30 (nominal), 120 (short-bunch)      & A        \\
        Twiss parameter $\beta_x$     & 1.4   & m  \\
        Twiss parameter $\beta_y$     & 0.24   & m  \\
        Twiss parameter $\alpha_x$     & 0.47   & rad  \\
        Twiss parameter $\alpha_y$     & 0  & rad \\
       \textbf{Undulator} &&   \\ 
        Period $\lambda_u$      & 2.3       & cm     \\ 
        Number of periods $N_u$   & 47     &      \\ 
        Undulator parameter $K$  & 1.2        &   \\ 
        Radiation wavelength $\lambda_r$      & 3.2     & $\mu$m\\
        \textbf{Cavity} &&   \\ 
        Length      & 2.0       & m    \\ 
        Mirrors radius of curvature  & 1.3    &    m  \\ 
        Cavity loss  & 7        &   $\%$\\ 
       \bottomrule
   \end{tabular}
   \label{tab:param}
\end{table}

\begin{figure*}[!tbh]
   \centering
   \includegraphics*[width=\textwidth]{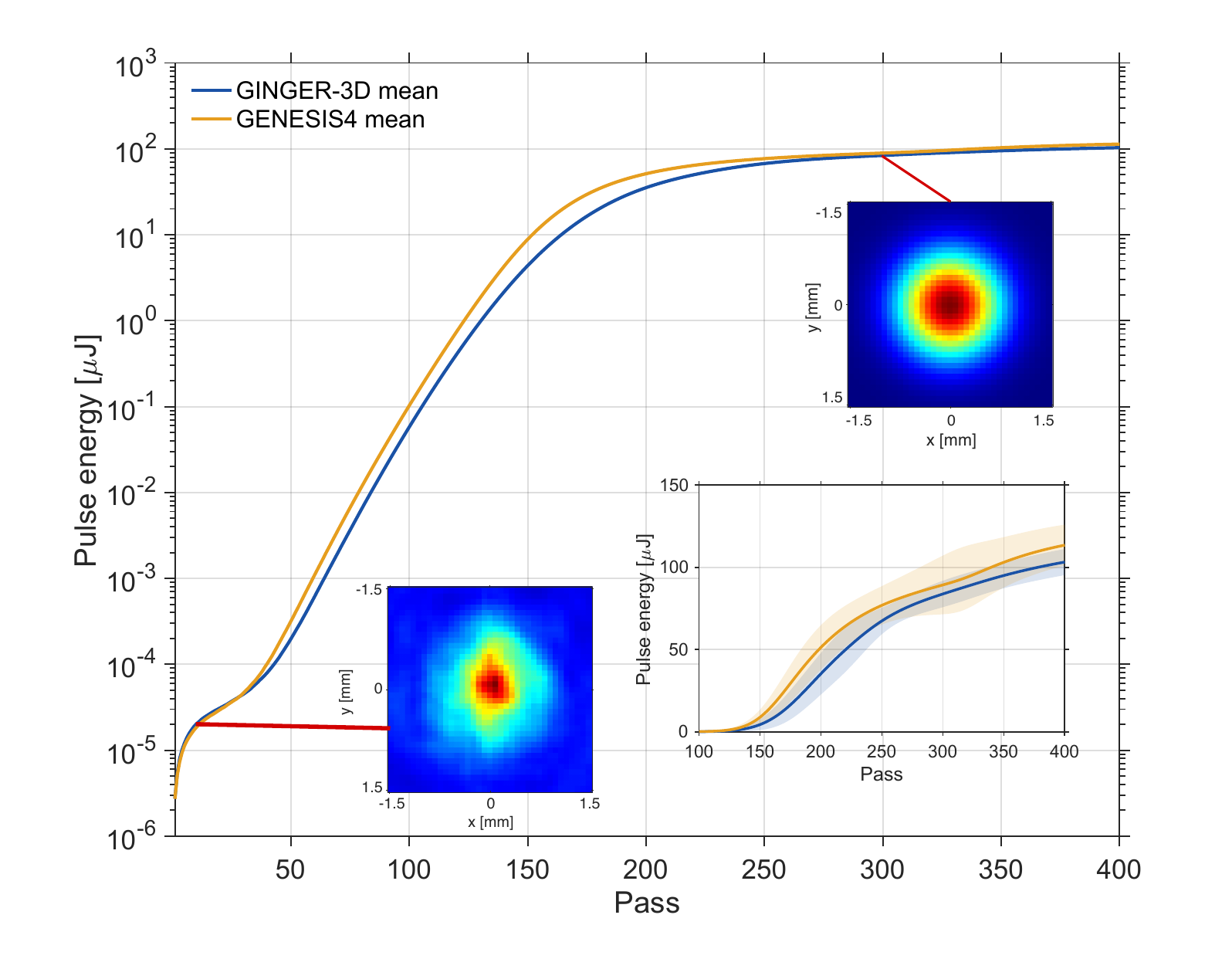}
   \caption{Radiation pulse energy as a function of oscillator pass number, averaged over 10 runs with different shot-noise seeds using the nominal beam and machine parameters, and benchmarked against the averaged \texttt{GENESIS4} simulation result. The inset shows the mean $\pm$ standard deviation as a shaded band on a linear scale. We show the integrated transverse radiation profile at passes 10 and 300 from one representative run among the 10 \texttt{GINGER-3D} simulations. Note that the colorbars are normalized to have a maximum of 1 for both profiles.}
   \label{fig:power_pass_2ps}
\end{figure*}





\begin{figure*}[t]
    \centering

    \begin{subfigure}[t]{0.48\textwidth}
        \centering
        \includegraphics[width=\linewidth]{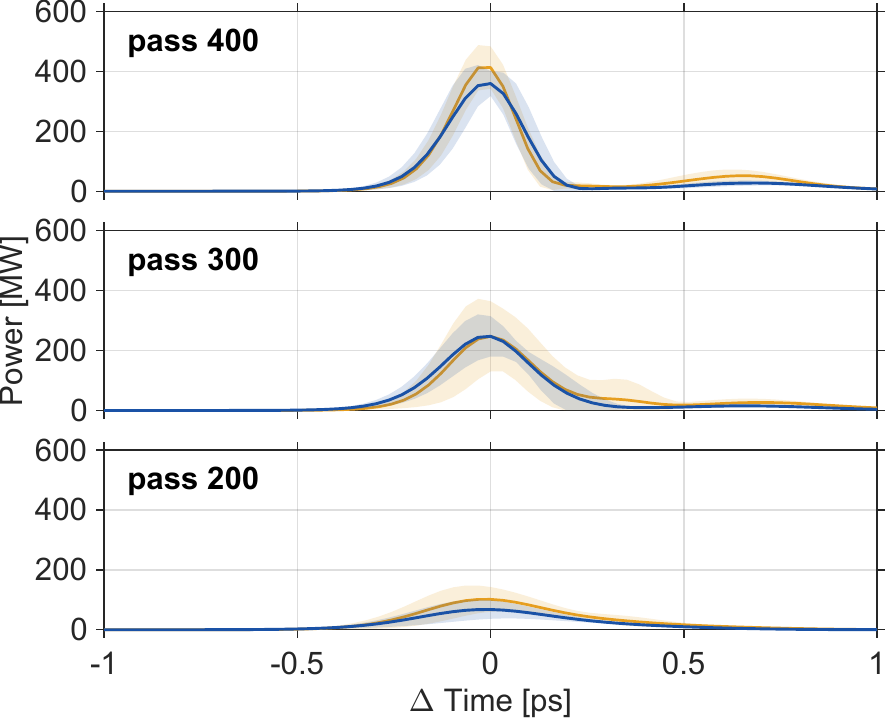}
    \end{subfigure}
    \hfill
    \begin{subfigure}[t]{0.48\textwidth}
        \centering
        \includegraphics[width=\linewidth]{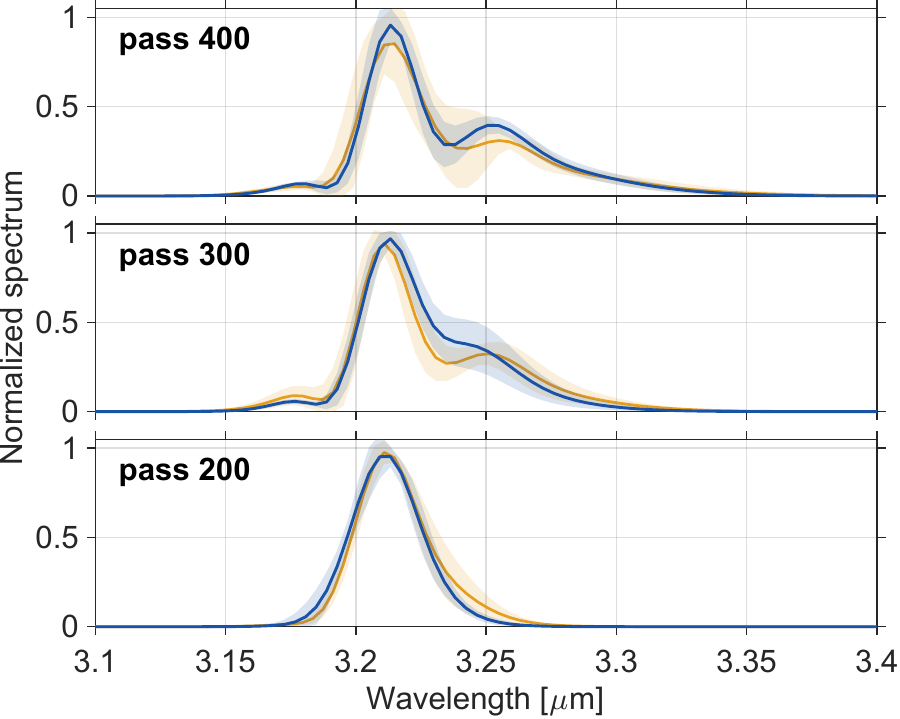}
    \end{subfigure}

    \caption{Comparison of \texttt{GENESIS4} and \texttt{GINGER-3D} time-dependent oscillator simulations. The left panel shows temporal pulse profiles at representative passes. The averaged pulse peaks are aligned to compare profile shapes. The right panel shows the corresponding averaged spectra. \texttt{GINGER-3D} results are shown in blue, and \texttt{GENESIS4} results are shown in yellow. Solid lines represent the average of the 10 runs, while the shaded bands indicate mean $\pm$ standard deviation.}
    \label{fig:benchmark_profile}
\end{figure*}


For the FEL interaction, we use the newly developed \texttt{GINGER-3D}, which offers radiation propagation capabilities similar to those of \texttt{GENESIS}, but is expected to be faster due to its Fortran 2003/2008 code base, in contrast to the C++ implementation of \texttt{GENESIS}. \texttt{GINGER-3D} also outputs data in HDF5 format, enabling seamless post-processing with standard tools such as \texttt{Python} and \texttt{MATLAB}.

To model radiation propagation outside the undulator, we have implemented a separate propagation module employing Fourier-optics-based free-space diffraction and mirror boundary conditions.
After each undulator pass, the output field from \texttt{GINGER-3D} is passed to this code, which simulates free-space propagation and reflection from the cavity mirrors. In the Fourier optics model, each mirror is approximated as a focusing lens with a focal length equal to half of its radius of curvature. 
The pulse propagation code is modular and allows convenient modification of cavity configurations, including asymmetric mirror reflectivity and the insertion of Michelson and Fox-Smith resonators~\cite{szarmes2002michelson1,szarmes2002michelson2}, to explore potential 3D effects in FEL oscillators. The propagated field is then re-injected into \texttt{GINGER-3D} for the next round-trip simulation. The \texttt{Matlab} and \texttt{Python} code package for pulse propagation can be found on~\cite{weinberg2025github}.

The FEL system at UH Mānoa consists of a thermionic microwave gun that produces electrons with macropulses of 4-5~$\mu$s duration at a 4~Hz repetition rate. Each macropulse contains micropulses at 2.856~GHz, set by the RF system. The linear accelerator (linac) accelerates the beam up to 45~MeV. At the FEL cavity entrance, the electron bunch can reach a full width at half maximum (FWHM) duration around 2~ps~\cite{niknejadi2019free}. The FEL uses a Mark V planar undulator and an optical cavity tuned to operate near 3~$\mu$m~\cite{benson1990review}.

Table \ref{tab:param} summarizes the machine settings and beam parameters. Because the machine is not operating at the time of writing, the machine settings and nominal beam parameters are taken from past measurements~\cite{benson1988status, benson1990review,niknejadi2019free}. 
In the vertical plane, the beam is initialized at a waist at the undulator entrance using the previously reported Twiss parameters, which provide a near-matched vertical envelope through the undulator.
In the horizontal plane, the waist is assumed to occur at the center of the undulator, with the horizontal beam size matched to the radiation beam size. The Rayleigh range $Z_R$ of the radiation is approximately half of the undulator length, ensuring strong overlap with the electron beam. Therefore, using $\sigma_x\sim\sigma_r=\sqrt{\frac{\lambda_r}{4\pi}Z_R}$, we set $\beta_x$ by $\beta_x=\sigma_x^2/\epsilon_x$, where $\epsilon_x$ is the geometric emittance.

The transverse radiation field is discretized on a 129x129
grid with spacing 80~$\mu$m, corresponding to a total transverse field of view of 1~cm by 1~cm.
Tests with finer transverse grids produced no appreciable change in the results, indicating that the chosen resolution is sufficient for the present simulations. The longitudinal temporal discretization is 33~fs per slice.

Figure \ref{fig:power_pass_2ps} shows the radiation pulse energy as a function of the number of round-trips through the cavity, averaged over 10 runs with different shot-noise seeds, using the nominal beam and machine parameters. The averaged radiation energy reaches microjoule level around 130 passes and reaches saturation around 200-300 passes. We limit the maximum pass number to 400 in the numerical simulations because it corresponds to about 5~$\mu$s of macropulse, the longest macropulse duration delivered at the UH Mānoa linac, limited by the pulse forming network and back-bombardment effect on the cathode~\cite{kowalczyk2013measurement}. Inset plots show the integrated transverse profile of the radiation pulse for one representative \texttt{GINGER-3D} run at pass 10 and pass 300, respectively, highlighting the spatial structure captured by our 3D simulation framework.


To validate the \texttt{GINGER-3D} simulations, we perform benchmark studies against \texttt{GENESIS4} in two stages. First, using the beam and machine parameters listed in Table~\ref{tab:param}, we carry out steady-state simulations in both codes and scan the carrier wavelength with an input seed laser. Both codes consistently produce the maximum gain near 3.2~$\mu$m. We therefore use 3.2~$\mu$m as the carrier wavelength for the subsequent time-dependent FEL oscillator simulations over the full 400 passes. 

In the time-dependent benchmark, \texttt{GENESIS4} is run with ten independent random initial seeds. The transverse grid size is $125~\mu\mathrm{m}$ with a $129\times129$ grid, selected from a first-pass start-up test, balancing numerical convergence and computational cost, and the temporal slice spacing is $32~\mathrm{fs}$, close to the $33~\mathrm{fs}$ slice used in \texttt{GINGER-3D}. 
The resulting FEL power evolution shows good agreement with the \texttt{GINGER-3D} result (Fig.~\ref{fig:power_pass_2ps}).
After the first pass, the two codes give nearly identical pulse energies. \texttt{GINGER-3D} gives an average pulse energy of $2.74\times10^{-6}~\mu\mathrm{J}$ with negligible variation over 10 runs, while \texttt{GENESIS4} gives $(2.71\pm0.06)\times10^{-6}~\mu\mathrm{J}$ over 10 seeds, corresponding to a relative difference of $0.86\%$. After the final pass, \texttt{GINGER-3D} gives $103.2\pm7.9~\mu\mathrm{J}$ over 10 runs, while \texttt{GENESIS4} gives $113.5\pm12.5~\mu\mathrm{J}$ over 10 seeds. The final-pass \texttt{GENESIS4} mean is about $10.0\%$ higher than the \texttt{GINGER-3D} mean, while remaining within the run-to-run fluctuations.

We further compare the temporal pulse profile and radiation spectrum at representative passes, including passes 200, 300, and 400, as shown in Fig.~\ref{fig:benchmark_profile}. The different initial seeds lead to different longitudinal mode evolution during the earlier passes, but by the last pass they exhibit a similar pulse profile.
The spectra from the two codes also show good agreement. As the pass number increases, the radiation grows near the central wavelength around $3.21~\mu\mathrm{m}$ and develops sidebands around $3.18~\mu\mathrm{m}$ and $3.25~\mu\mathrm{m}$, consistent with temporal spiking and synchrotron oscillations of trapped electrons.

\section{\label{sec:desync}Desynchronization Simulation}

\begin{figure}[!tbh]
    \centering
    \begin{subfigure}[b]{0.45\textwidth}
        \includegraphics[height=5cm]{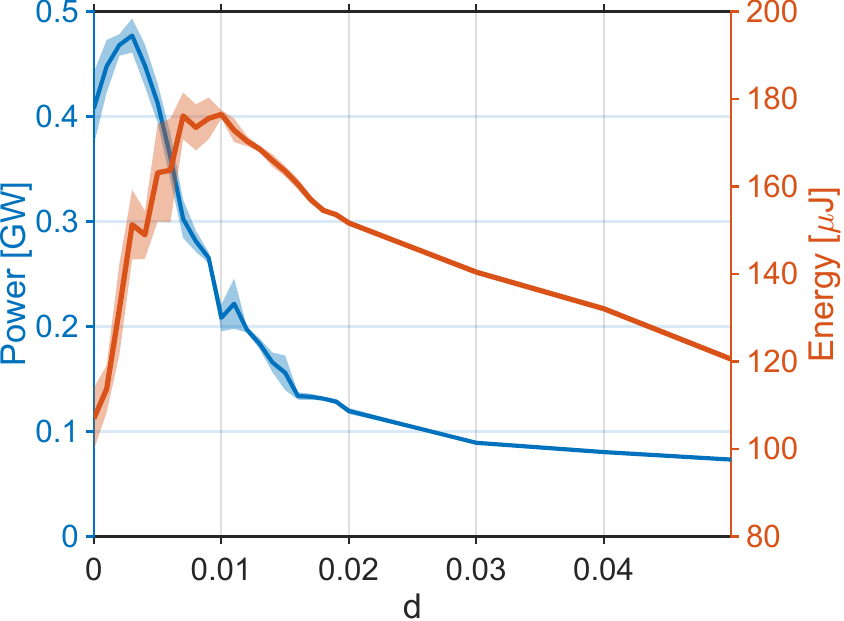}
        \caption{}
    \end{subfigure}
    \begin{subfigure}[b]{0.45\textwidth}
        \includegraphics[height=5cm]{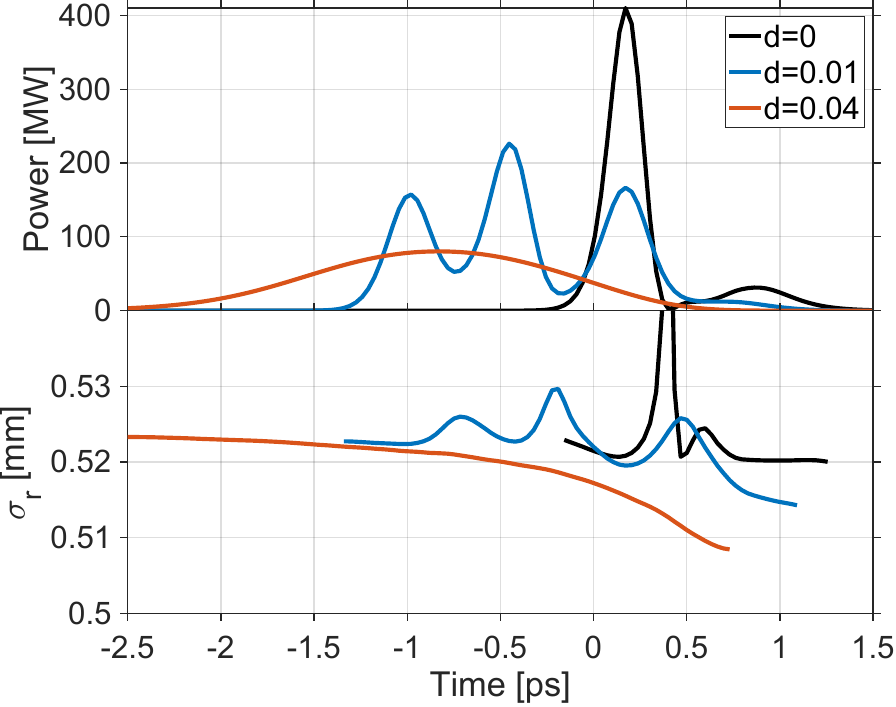}
        \caption{}
    \end{subfigure}
 \vspace{0.5cm} 
    \begin{subfigure}[b]{0.45\textwidth}
        \includegraphics[width=\linewidth]{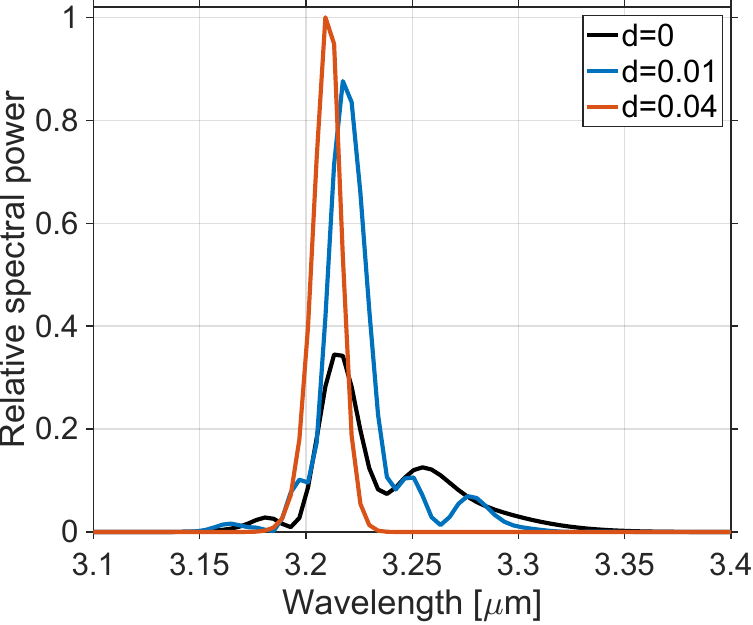}
         \caption{}
    \end{subfigure}
    \begin{subfigure}[b]{0.45\textwidth}
        \includegraphics[width=\linewidth]{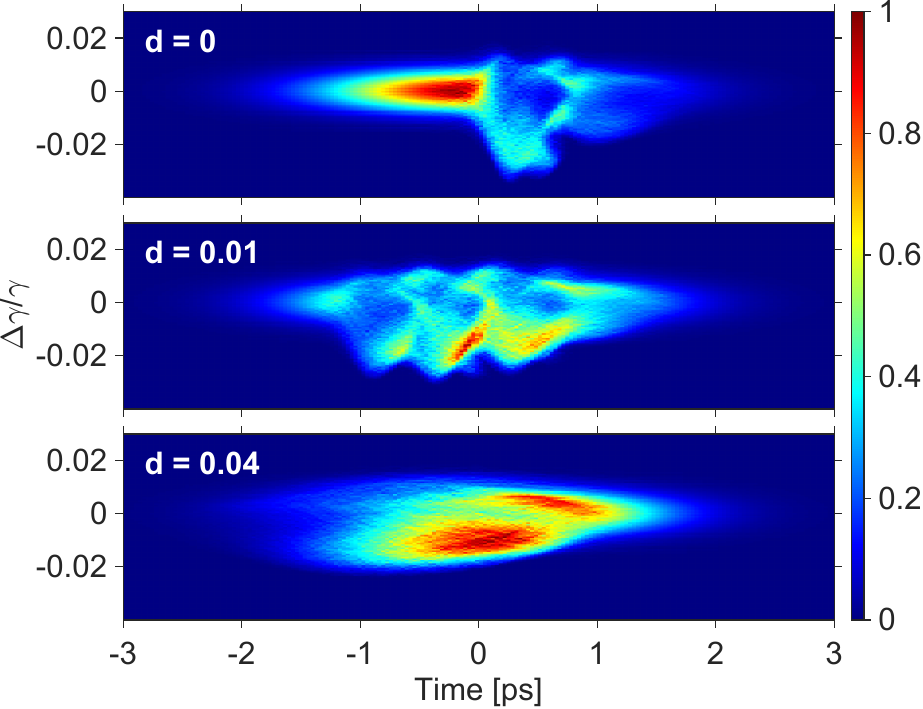}
        \caption{}
    \end{subfigure}
    \caption{2~ps bunch length: (a) Peak power (blue) and pulse energy (red) at oscillator pass 400 as a function of varying desynchronization values $d$. Shaded bands indicate mean~$\pm$~standard deviation over 5 runs with different shot-noise seeds. Panels (b)--(d) show results from one representative shot-noise seed selected from the ensemble used in panel (a). (b) Pulse profiles for three different values of $d$. The corresponding radiation spot sizes $\sigma_r$ are shown underneath the pulse profiles. (c) Normalized radiation spectrum at pass 400. (d) Longitudinal electron phase space after the final pass. Colorbar is normalized to have a maximum of 1. Time axes in (b) and (d) have been offset to position the electron beam at $t=0$.}
    \label{fig:profiles_2ps}
\end{figure}

As discussed in Sec.~\ref{sec:intro}, cavity desynchronization is introduced to mitigate the laser lethargy effect in FEL oscillators. We define the desynchronization value, $d=2\Delta L/S$, where $S = N_u \lambda_r$ is the slippage length, and $\Delta L$ is illustrated in Fig.~\ref{fig:schematics}. 
In our simulation, we implement this by shifting the radiation field in \texttt{GINGER-3D} by $2\Delta L$ during the free-space propagation before re-injection into the undulator.  
We scan a range of desynchronization values 
corresponding to round-trip shortening from a fraction of the radiation wavelength to several radiation wavelengths. For each $d$ value, we repeat the simulation five times with different shot-noise seeds, and the results are shown in Fig.~\ref{fig:profiles_2ps}(a). 

Focusing on one representative run, Fig.~\ref{fig:profiles_2ps}(b) shows the radiation pulse profiles after 400 passes for several desynchronization values. 
At perfect synchronization and small desynchronization (round-trip shortening of 1.5~$\mu$m), we observe the formation of spiking, as reported in~\cite{warren1986spiking, richman1989first}, and the latter case enhances the radiation pulse energy by 63\% on average.
This enhancement results from improved temporal overlap between the radiation pulse and electron bunch throughout the amplification process. At large desynchronization (round-trip shortening of 6~$\mu$m) the radiation pulse evolves into a smooth and stable near-Gaussian profile after around pass 200-300. The transition from spiking to stabilized behavior is also evident in the spectral domain, as shown in Fig.~\ref{fig:profiles_2ps}(c), where at large $d$ the spectrum resembles a smooth Gaussian, consistent with the predictions in~\cite{kim1991stability}.

When spiking occurs, the radiation pulse also undergoes temporal shortening. The FWHM of the leading spike is 0.2~ps for perfect synchronization and is similar for small desynchronization, comparable with the cooperation length of the FEL, whereas the FWHM pulse width is 1.5~ps for large desynchronization, where spiking is suppressed. This behavior is consistent with the saturation dynamics. Near synchronization, the radiation repeatedly overlaps with the same high-gain portion of the electron bunch, allowing a localized temporal region to grow rapidly, trap electrons in the ponderomotive potential, and evolve into a shortened high-intensity spike. By contrast, large desynchronization reduces the repeated overlap of the optical pulse with the same temporal slices of the electron bunch, suppressing the formation of narrow spikes.


With the 3D simulation capability, we examine the transverse radiation spot size for individual temporal slices after the spikes have formed. We quantify the spot size from the second moments of the transverse intensity distribution for each time slice, $\sigma_{rx}^2=\int\frac{x^2I(x,y)dxdy}{I(x,y)dxdy}$, $\sigma_{ry}^2=\int\frac{y^2I(x,y)dxdy}{I(x,y)dxdy}$, where $I(x,y)$ is the transverse radiation intensity at a given longitudinal time slice. The average spot size, $\sigma_r$, is then defined as the average of $\sigma_{rx}$ and $\sigma_{ry}$. The averaged spot size at the final pass is shown in Fig.~\ref{fig:profiles_2ps} (b). Optical guiding is most apparent in the spiking regime because the high-intensity spikes produce localized regions of stronger electron microbunching and gain, which reduce the transverse spot size near the power peaks. Between spikes, the radiation intensity is lower and the electron microbunching is reduced, so the gain-guiding contribution is weaker. In the large desynchronization case, the pulse is smoother and wider in time, so the interaction is distributed over a broader longitudinal region. The transverse mode then reflects the averaged gain evolution rather than localized guiding around individual spikes.

We examine the electron phase-space evolution after the final pass for small and large desynchronization, as shown in Fig.~\ref{fig:profiles_2ps}(d). The radiation spike is accompanied by strong energy modulation of the electron bunch, indicating localized energy exchange between the electrons and the radiation field. These longitudinally modulated regions persist near saturation, consistent with electron trapping and synchrotron motion in the ponderomotive potential (see Supplemental Material [URL will be inserted by publisher] SM2psd0p0105062026.mp4).


\begin{figure*}[!tbh]
   \centering
   \includegraphics[width=\textwidth]{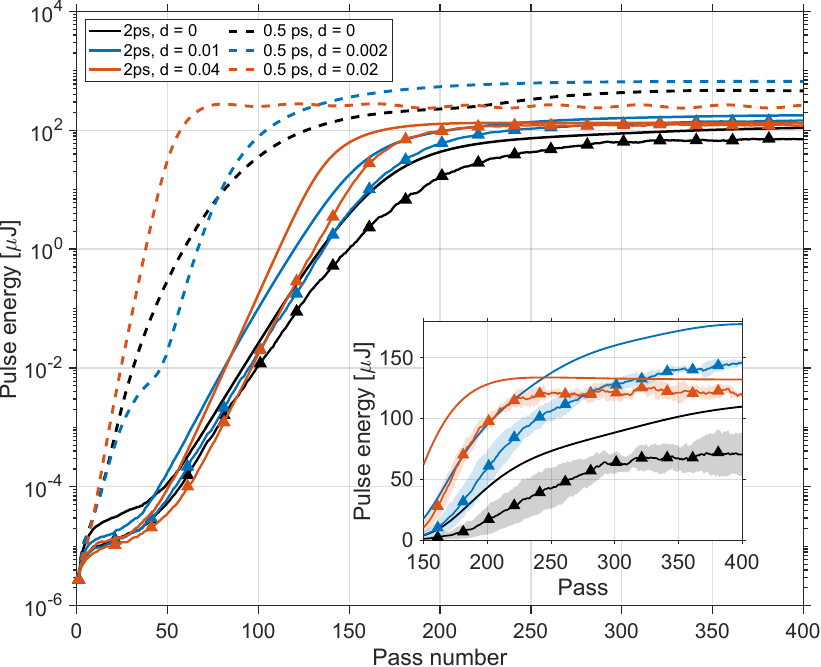}
   \caption{Pulse energy as a function of oscillator passes. Solid curves show 2~ps cases, and dashed curves show the no-jitter 0.5~ps cases. Solid curves with triangles show the average over five jitter runs for the 2~ps cases. The inset shows the same 2~ps jitter comparison on a linear scale, with the shaded bands indicating mean $\pm$ standard deviation. Colors denote the cavity desynchronization parameter $d$.}
   \label{fig:power_pass}
\end{figure*}

Electron-bunch arrival-time jitter can influence the effectiveness of cavity desynchronization in compensating for laser lethargy, since it randomizes the relative timing between the electron bunch and the radiation pulse. To quantify this effect, we carry out additional simulations for the 2~ps bunch case at the three desynchronization values in Fig.~\ref{fig:profiles_2ps}(b). The bunch arrival time is varied randomly pass by pass according to a Gaussian distribution with an rms value 0.4~ps, corresponding to the 0.4$^\circ$ RF phase jitter reported for the UH S-band linac~\cite{lumpkin2013initial,hadmack2013electron} under nominal operating conditions. We repeat the simulation five times with different random jitters from the same 0.4~ps rms distribution, using the same shot-noise seeds, and the averaged results are shown in Fig.~\ref{fig:power_pass}. 
Timing jitter reduces the radiation energy most strongly near synchronization. At $d=0$, the final pulse energy drops from 109.8 to 70.3~$\mu$J, a 36.0\% degradation. The effect becomes much smaller with increasing desynchronization, 18.2\% degradation at $d=0.01$ and 9.9\% at $d=0.04$. Near synchronization, even though the oscillator retains higher radiation power, the strong dependence on electron-bunch and radiation-pulse overlap makes it more sensitive to arrival-time jitter. In contrast, large desynchronization suppresses spiking and sideband growth while favoring a smoother and broader longitudinal mode, making the oscillator less sensitive to pass-to-pass timing fluctuations.

\section{\label{sec:shortbunch}Short-Bunch Operation with Desynchronization}

\begin{figure*}[!tbh]
    \centering
     \begin{subfigure}[b]{0.45\textwidth}
        \includegraphics[height=5cm]{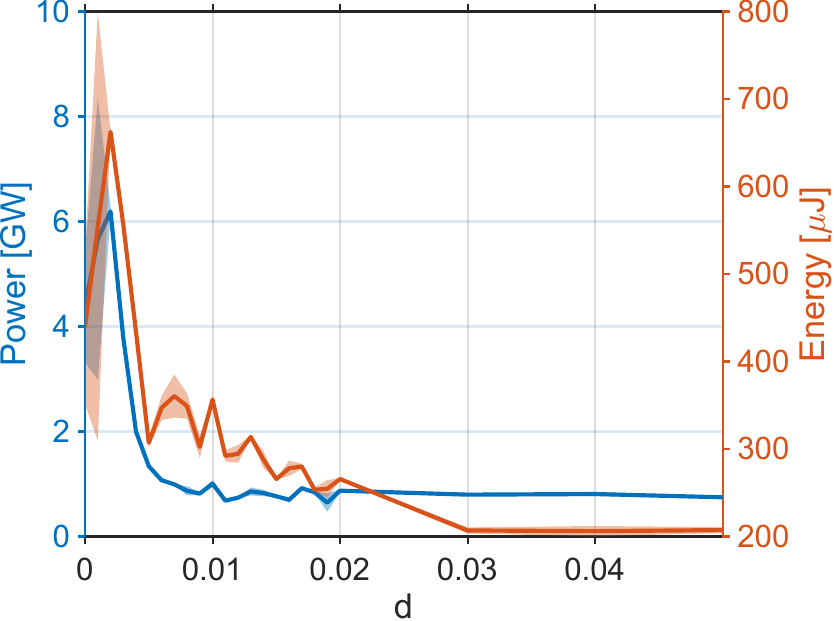}
        \caption{}
    \end{subfigure}
    \begin{subfigure}[b]{0.45\textwidth}
        \includegraphics[height=5cm]{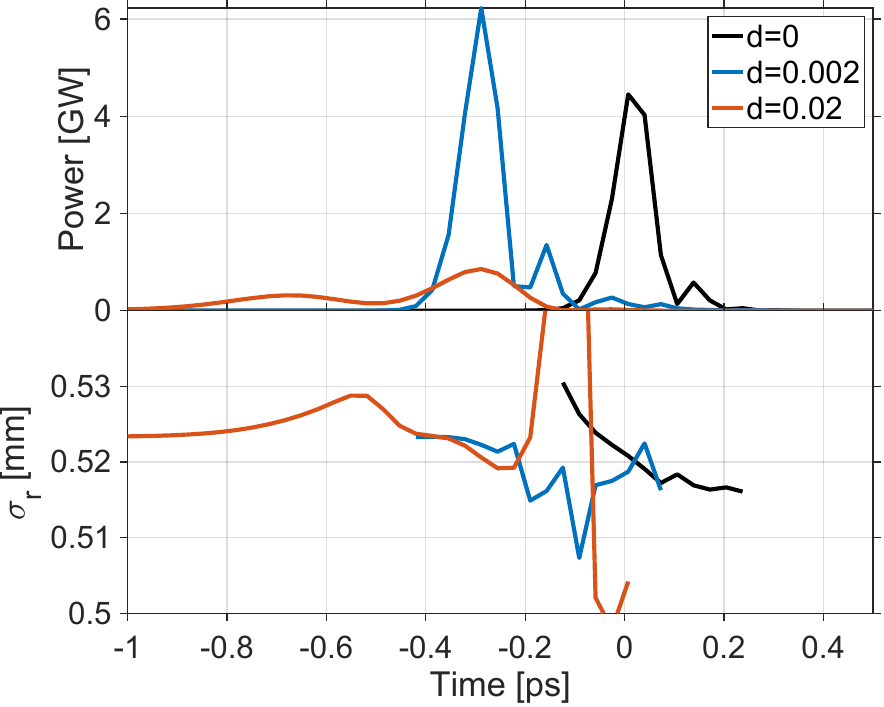}
        \caption{}
    \end{subfigure}
 \vspace{0.5cm} 
    \begin{subfigure}[b]{0.45\textwidth}
        \includegraphics[width=\linewidth]{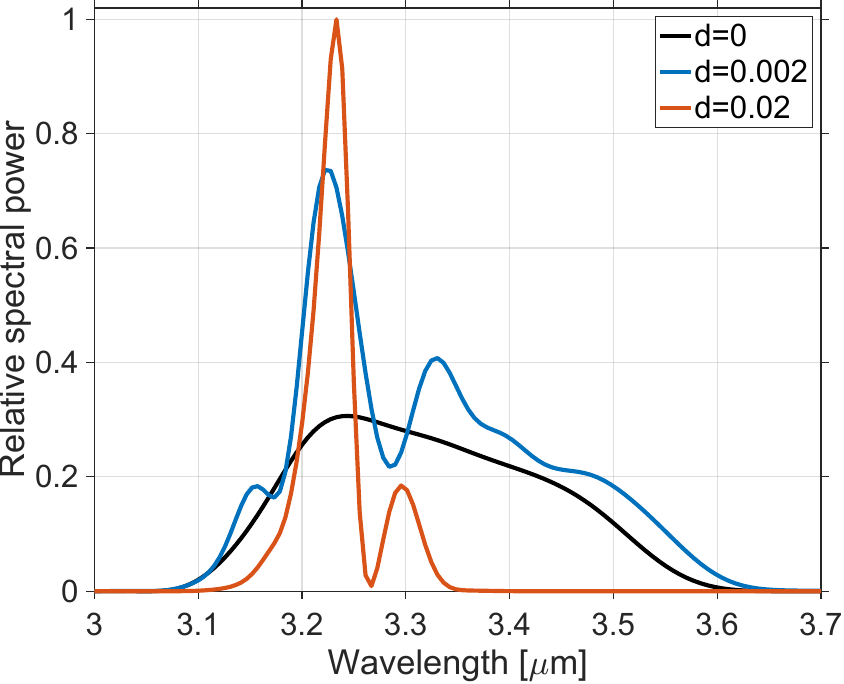}
        \caption{}
    \end{subfigure}
    \begin{subfigure}[b]{0.45\textwidth}
        \includegraphics[width=\linewidth]{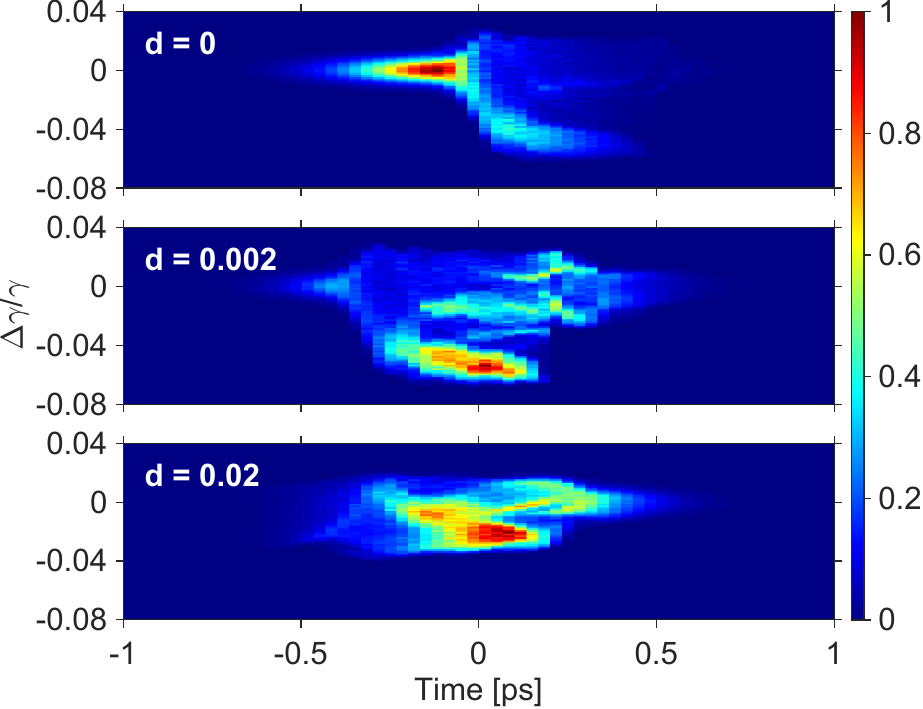}
        \caption{}
    \end{subfigure}
    \caption{0.5~ps bunch length: (a) Peak power (blue) and pulse energy (red) at oscillator pass 400 as a function of varying desynchronization values $d$. Shaded bands indicate mean~$\pm$~standard deviation over 5 runs with different shot-noise seeds. Panels (b)--(d) show results from one representative shot-noise seed selected from the ensemble used in panel (a). (b) Pulse profiles for three different values of $d$. The corresponding radiation spot sizes $\sigma_r$ are shown underneath the pulse profiles. (c) Normalized radiation spectrum at pass 400. (d) Longitudinal electron phase space after the final pass. Colorbar is normalized to have a maximum of 1. Time axes in (b) and (d) have been offset to position the electron beam at $t=0$.}
    \label{fig:profiles_0.5ps}
\end{figure*}

To go beyond the nominal operation parameters at UH, we simulate FEL oscillator operation using a shorter electron bunch with a FWHM duration of 0.5~ps, which is comparable to the slippage length, and roughly two cooperation lengths.

Note that we set the simulation time slice to be about three times the resonant radiation period. This choice is consistent with the slowly-varying envelope approximation employed by \texttt{GINGER-3D}, in which the field envelope varies on timescales longer than the radiation period.
In the short-bunch case, the FEL reaches saturation significantly faster than in the nominal 2~ps case, as shown in Fig.~\ref{fig:power_pass}. Even in perfect synchronization, the peak power increases from 410~MW in the 2~ps case to 4.45~GW in the 0.5~ps case. 

We repeat the desynchronization scan described in Sec.~\ref{sec:desync} to study its effect on the radiation pulse evolution. 
In Fig.~\ref{fig:profiles_0.5ps}(a), we observe a relatively large shot-to-shot variation at small desynchronization, where the result is more sensitive to the initial shot-noise seed. This suggests that, for such fine desynchronization, the oscillator may select different longitudinal mode structures, leading to more or less favorable temporal overlap with the electron bunch.

For $d\ge0.002$, the imposed cavity shortening becomes the dominant effect, and the radiation pulse is consistently shifted toward the head of the bunch, leading to improved overlap and the observed enhancement in output power. We note that this optimal desynchronization is smaller than the $d=0.01$ case for the 2~ps bunch. Because the short-bunch saturates more quickly, the simulation after 400 passes represents a regime where gain is slowed and the lethargy effect has already relaxed. This result underscores the importance of dynamic desynchronization for short-bunch operation, which we will explore in future work. 

We also see the dependence of the peak power and the pulse energy on the desynchronization value is similar. This is because in the short-bunch case at small desynchronization values, the bunch length only allows for one dominant spike followed by a weak trailing spike to develop instead of the multiple spikes shown in Fig.~\ref{fig:profiles_2ps}(b), making both the peak power and pulse energy similarly sensitive to the desynchronization values.

Here, the optical-guiding signature is weak at $d=0$ and $d=0.002$ because the radiation pulse remains very short, and the radiation slips across a large fraction of the electron bunch within one pass. As a result, the transverse mode is coupled to a rapidly changing longitudinal overlap rather than a clear narrowing of the transverse spot size at each power peak. At larger desynchronization, the 0.5~ps case also differs from the stabilized behavior observed for the 2~ps bunch. For the 0.5~ps bunch at $d=0.02$, the pulse broadens and develops multiple temporal peaks, with corresponding transverse spot-size variation. This indicates that optical guiding begins to reappear, resembling the behavior of the 2~ps bunch at $d=0.01$.

\section{\label{sec:conclusions} Conclusions}
In conclusion, we have developed and demonstrated a comprehensive simulation framework for modeling the FEL oscillator at the University of Hawai`i at Mānoa. 
The simulations reveal fully three-dimensional, time-dependent oscillator dynamics and electron beam phase-space evolution that are not captured in one-dimensional models. 
For the nominal operating parameters, our results show that modest cavity desynchronization can enhance the output radiation power, but the resulting spiking behavior in the longitudinal modes makes the oscillator sensitive to machine timing jitter. At larger desynchronization, temporal spiking and spectral sidebands are suppressed, leading to smoother pulse evolution and improved robustness against timing jitter, although at the expense of reduced final radiation power. 
We also examine a short-bunch regime and find enhanced radiation power and earlier saturation compared with the nominal case, while dependence on desynchronization shows distinctly different behaviors compared with the nominal case due to the short bunch length.
Together, these results suggest practical operating regimes for pulse control and open up new possibilities for high-peak-power and ultrafast pulse generation at UH Mānoa. This work also establishes a flexible numerical platform for future studies of dynamic desynchronization, optimization, and control.
Looking forward, current efforts on reviving the facility will make it possible to experimentally verify the results of our simulations. Furthermore, this simulation framework enables detailed studies of advanced optical configurations, such as interferometric setups for imposing phase coherence on the output pulses, as well as systematic investigations into cavity length tuning for maximizing FEL output power, extraction efficiency, and bunch length optimization to support future beamline upgrades.

\begin{acknowledgments}
The authors acknowledge William Fawley and Niels Bidault for helpful discussions. 
Work on the \texttt{GINGER-3D} code is supported by the U.S. Department of Energy under Contract No. DE-AC02-76SF00515 to SLAC and in part by Sincrotrone Trieste.
This work is supported by the U.S. Department of Energy, Office of Science, Office of Basic Energy Sciences and Office of Accelerator Research \& Development and Production, under Contract No. DE-SC0025583.
\end{acknowledgments}

\bibliography{mainbib}

\end{document}